\documentclass[12pt,a4paper]{article}
\usepackage[cm]{fullpage}
\usepackage[affil-it]{authblk}
\usepackage{xcolor}
\usepackage[breaklinks=true,colorlinks,citecolor=blue,linkcolor=blue,urlcolor=blue]{hyperref}
\usepackage{amsmath}
\usepackage{amsfonts}
\usepackage{amssymb}
\usepackage{bm}
\usepackage{physics}
\usepackage[pdftex]{graphicx}

\begin{document}

\title{A regional implementation of a mixed finite-element, semi-implicit dynamical core}

\author[1,*]{Christine Johnson}
\author[1]{Ben Shipway}
\author[1]{Thomas Melvin}
\author[1]{Thomas Bendall}
\author[1]{James Kent}
\author[1]{Ian Boutle}
\author[1]{Alex Brown}
\author[1]{Mohamed Zerroukat}
\author[1]{Benjamin Buchenau}
\author[1]{Nigel Wood}

\affil[1]{Met Office, Exeter, UK}
\affil[*]{Corresponding Author: christine.johnson@metoffice.gov.uk}

\maketitle

\let\thefootnote\relax\footnotetext{This pre-print has been submitted to the Quarterly Journal of the Royal Meteorological Society.}
 
\begin{abstract}

This paper explores how to adapt a new dynamical core to enable its use in one-way nested regional weather and climate models, where lateral boundary conditions (LBCs) are provided by a lower-resolution driving model. The dynamical core has recently been developed by the Met Office and uses an iterated-semi-implicit time discretisation and mixed finite-element spatial discretisation.

The essential part of the adaptation is the addition of the LBCs to the right-hand-side of the linear system which solves for pressure and momentum simultaneously. The impacts on the associated Helmholtz preconditioner and multigrid techniques are also described.

The regional version of the dynamical core is validated through big-brother experiments based on idealised dynamical core tests. These experiments demonstrate that the subdomain results are consistent with those from the full domain, confirming the correct application of LBCs. Inconsistencies arise in cases where the LBCs are not perfect, but it is shown that the application of blending can be used to overcome these problems.
\\

\end{abstract}%

\section{Introduction}
In weather and climate forecasting, improved accuracy can be obtained by running the numerical model with finer resolution grids and more sophisticated representations of the physical processes. However, this needs ever more computing power. In recent years that increase in computing power has been achieved by changes to supercomputer architectures. Over the last decade there has been more focus on parallel processing, requiring numerical models to improve their scalability. And in the near future, supercomputers will have a greater diversity of hardware, requiring bespoke optimisations and programming languages.

To harness the power of these diverse supercomputer architectures, the Met Office has developed a new dynamical core, GungHo \cite[]{Melvin:2019, Melvin:2024}. This addresses the scalability bottleneck caused by the singularity at the poles on a global longitude-latitude grid by switching to a cubed-sphere mesh. Additionally, the numerical schemes for transport have been improved using new inherently conservative and consistent finite-volume schemes \cite[]{Bendall:2024}. This new dynamical core also uses a new infrastructure, known as LFRic \cite[]{Adams:2018}, which was created with a design philosophy of ‘Separation of Concerns', separating the science code from supercomputer optimisation code and enabling use on diverse hardwares.

The Met Office uses both global and regional models for weather and climate forecasting. Global models predict the evolution of large weather systems and climate signals over days to years, while regional models provide enhanced resolution and detail over smaller, limited-area regions, with the Met Office using a rotated-pole, lat-lon grid. These regional models form the basis of the Met Office’s operational high-resolution numerical weather predictions \cite[]{Tang:2013, Bush:2020, Bush:2023}, regional climate studies \cite[]{Kendon:2014}, convective-scale ensembles \cite[]{Hagelin:2017} and more recently sub-km, hectometric scale models \cite[]{Lean:2024}. Therefore, a regional version of the GungHo dynamical core is essential for our future forecasting capabilities.

The concept of regional models is not new, and various studies have examined how to handle lateral boundary conditions (LBCs). Early studies focused on finding appropriate inflow and outflow boundary conditions to give well-posed mathematical equations e.g. \cite[]{Oliger:1978}. Whilst this approach has been implemented in simple models e.g. \cite[]{McDonald:2003}, it is not straightforward for multi-level, full-models with inconsistencies in the boundary data. The alternative approach known as relaxation \cite[]{Davies:1976} or blending \cite[]{Davies:2014} is commonly used. The method, also adopted here, allows the interior solution to diverge from the driving model while ensuring that the solution matches at the boundaries by blending LBC data with the interior solution.

This paper describes the limited area model version of the GungHo dynamical core. We begin by describing the approach to creating a regional model — a general philosophy applicable to generating a regional model for any dynamical core. We then describe the continuous equations and spatial and temporal discretisations that form the basis of the GungHo dynamical core and how these discretisations can be adapted for the regional model. This is followed by computational experiments based on idealized dynamical core tests.

\section{Philosophy of Approach}

The aim here is to use the regional model in a one-way nesting, where the regional model is nested in a low-resolution driving model and information is passed from the driving model to the regional model, but not vice versa. This is in contrast to other philosophies such as mesh refinement e.g. \cite[]{Zarzycki:2014}, mesh stretching e.g. \cite[]{Sergeev:2024}, and two-way nesting e.g. \cite[]{Zhang:1986} that are used elsewhere.

To enable this one-way nesting we first consider the situation where the regional model is able to exactly replicate the driving model over the limited area region. This is analogous to the big-brother and perfect-model experiments described in for example, \cite{Davies:2014} and \cite{deElia:02}, where the solution from the subdomain is compared with the solution from running the model over the entire domain, and the LBC data is derived from the full-domain run where the regional model has exactly the same resolution and numerical schemes as the driving model.

The big-brother approach can be summarised as:
\begin{enumerate}
\item Within the subdomain, the regional numerical model updates the values.
\item Outside the subdomain, the values are updated by copying data from a scenario where the numerical model has been run over the entire domain (with the same numerical schemes and spatial and temporal resolutions). This is the lateral boundary condition (LBC) data.
\item  The results should be identical (within the tolerance of the numerical implementation) to those obtained by using the numerical model to update values over the whole domain.
\end{enumerate}

The big-brother experiment is typically used to verify the validity of the regional model, but in this paper we extend the concept further by using points 1-3 to design the regional model. The aim is to design the regional model by adapting the full model equations to ensure that points 1-3 are satisfied.

This formulation can then be extended later, through the use of blending, to allow for the case where the regional model is different to the model used to create the LBC data, for example with a lower-resolution driving model.

\section{Dynamical core for a global domain}

To explain how the dynamical core for the regional domain is derived, we first describe the dynamical core for the global domain. This is essentially the same as the formulation presented in \cite{Melvin:2024}, but additionally using the eliminated density and potential temperature in the solver as described in \cite{Maynard:2020} and the SWIFT transport scheme of \cite{Bendall:2024}.

\subsection{Continuous Equations}
Central to the dynamical core is the solution of the Euler equations for a perfect gas in a rotating frame, with the equation of state,
 \begin{equation}
    \begin{aligned}
      \frac{\partial \mathbf{u} }{\partial t} & = - \mathbf{u} \cdot \grad \mathbf{u} - 2 \bm \Omega \times \mathbf{u} - \grad \Phi - c_p \theta \grad \Pi \\
      \frac{\partial \rho}{\partial t} & = - \grad \cdot ( \rho \mathbf{u}) \\
      \frac{ \partial \theta}{ \partial t} & = - \mathbf{u} \cdot \grad \theta \\
    \Pi ^ {\left( \frac{1 - \kappa}{\kappa} \right)} & = \frac{R}{p_0} \rho \theta
    \end{aligned}
  \end{equation}
where the prognostic variables are the velocity vector $\mathbf{u}$, density $\rho$, potential temperature $\theta$ and the Exner pressure $\Pi$. $\bm \Omega$ is the rotation vector, $\Phi$ is the gravitational geopotential, $c_p$ is the specific heat at constant pressure, $R$ is the gas constant per unit mass, $p_0$ is the reference pressure, and $\kappa\equiv R/c_p$. Boundary conditions of zero-mass flux are also applied at the top and bottom boundaries.

\subsection{Iterated semi-implicit time-discretisation}

An iterated semi-implicit scheme is used for the time-discretisation; to calculate the state at the next timestep $t^{n+1}$, then quantities at both the current ($t^n$) and next ($t^{n+1}$) are used, and the resulting equations are solved iteratively using a quasi-Newton method, such that  e.g. $\theta_{k+1}^{n+1} = \theta_{k}^{n+1} + \theta'$.  When this method is applied to the Euler equations then the following system needs to be solved for the increments, such that on each iteration $k+1$
\begin{equation}\label{eq:iterated}
\begin{aligned}
  \mathbf{u}' + \Delta t \mu \mathbf{z}(\mathbf{z} \cdot \mathbf{u}') + \tau_u \Delta t( 2 \bm \Omega \times \mathbf{u}') & \\
   + \tau_u \Delta t c_p (\theta' \grad \Pi^* + \theta^* \grad \Pi' ) & = \mathbf{r}_u \\
\rho' + \tau_\rho \Delta t \grad \cdot (\rho^* \mathbf{u}') & =  r_\rho \\
\theta' + \tau_\theta \Delta t (\mathbf{z} \cdot \mathbf{u}') \frac{\partial \theta^*}{\partial z} & = r_\theta \\
\frac{\Pi'}{\Pi^*} - \frac{\kappa}{1 - \kappa} \left( \frac{\rho'}{\rho^*} + \frac{\theta'}{\theta^*} \right) & = r_\Pi 
\end{aligned}
\end{equation}
where $\tau_{u,\rho,\theta}$ are relaxation parameters, $\mathbf{z}$ is the unit normal vector in the vertical direction, and $\mu$ is a damping profile term associated with implicit damping of the vertical component of velocity. Only the vertical component of the buoyancy coupling is retained in the potential-temperature equation. The (*) terms are the reference state values and $r_{u,\rho,\theta,\Pi}$ are the right-hand-side residual terms calculated on that iteration:
\begin{equation}\label{eq:rhs}
  \begin{aligned}
     \mathbf{r}_u & = \Delta t \left( \delta_t \mathbf{u} + \mu \mathbf{z}(\mathbf{z} \cdot \mathbf{u}) + \bm{ \mathcal{A}^A} \left( \mathbf{u}^p, \bar{\mathbf{u}}^{1/2} \right) + \bar{\bm S}^{\alpha} \right) \\
     r_\rho & = \Delta t \left( \delta_t \rho + \grad \cdot \mathcal{F} \left( \rho^p, \bar{\mathbf{u}}^{1/2} \right) \right)\\
     r_\theta & = \Delta t \left( \delta_t \theta + \mathcal{A}^C \left( \theta^p, \rho^p, \bar{\mathbf{u}}^{1/2} \right) \right) \\
     r_\Pi & = 1- \frac{p_0}{R \rho \theta} \Pi^{\frac{1-\kappa}{k}} \\
  \end{aligned}
\end{equation}

where
\begin{equation}
  \begin{aligned}
  \bm S & = -2 \bm \Omega \cross \mathbf{u} - \grad \Phi - c_p \theta \grad \Pi \\
  \delta_t F & \equiv \frac{F^{n+1} - F^n}{\Delta t} \\
  \bar{F}^{\alpha} & \equiv \alpha F^{n+1} + ( 1- \alpha) F^n
\end{aligned}
\end{equation}
and $\alpha$ is a scalar, off-centring parameter. $\mathcal{F} \left( \rho^p, \bar{\mathbf{u}}^{1/2} \right)$ is the time-averaged flux, $\bm{ \mathcal{A}}^A \left( \mathbf{v}^p, \bar{\mathbf{u}}^{1/2} \right)$ is the time-averaged advection tendency of a vector $\mathbf{v}$ using an advective equation form, $\mathcal{A}^C \left( \theta^p, \rho^p, \bar{\mathbf{u}}^{1/2} \right)$ is the time-averaged advection tendency of a scalar $\theta$ using a consistent equation form and $p$ denotes an intermediate `predictor' state (see \cite{Kent:2023}). Please see \cite{Bendall:2024} for a discussion of advective, consistent and conservative forms of equations.

As in \cite{Wood:2014} and \cite{Melvin:2024}, the Newton iterations are split into $n_o$ outer loops and $n_i$ inner loops, with the transport terms only being updated in the outer loops. This gives a total of $n = n_o \times n_i$ iterations, with $n_o=2$ and $n_i=2$ being used here.

\subsection{Finite-element spatial discretisation}

A mixed finite-element spatial discretisation is used;  $\mathbf{u}$, $\theta$, $\rho$ and $\Pi$ are represented as the linear combination of prescribed basis functions. For the increments, this is:

\begin{equation}\label{eq:fes}
  \begin{aligned}
  \mathbf{u}'(x) & = \sum_i \xi_u^i \mathbf{v}_i (x) \\
  \theta'(x) & = \sum_i \xi_\theta^i \omega_i(x) \\
  \rho'(x) & = \sum_i \xi_\rho^i \sigma_i(x) \\
  \Pi'(x) & = \sum_i \xi_\Pi^i \sigma_i(x) \\
  \end{aligned}
\end{equation}
where for example $\xi_u^i$ is the scalar coefficient associated with the basis function $\mathbf{v}_i (x)$.

\begin{itemize}
  \item $ \sigma_i \in \mathbb{W}_3$ - the discontinuous-Galerkin space of scalar functions (volumes).
  \item $\mathbf{v}_i \in \mathbb{W}_2$ - the Raviart-Thomas space of vector functions (fluxes), with continuous normal components in both the horizontal and vertical, and such that $\grad \cdot \mathbf{v}_i \in \mathbb{W}_3$.
  \item $\omega_i \in \mathbb{W}_\theta $ - the space of scalar functions based on the vertical part of $\mathbb{W}_2$, that are continuous in the vertical but discontinuous in the horizontal.
\end{itemize}

To deal with solving the equations on the sphere with any horizontally unstructured mesh, a coordinate transformation $\phi$ is used to map from a unit-cube computational space $\hat{\chi}$ to the physical space $\chi$, such that $\chi = \phi(\hat{\chi})$. See \cite{Melvin:2024} for further details.

To discretise the continuous equations using finite elements, they must first be transformed into the weak form by multiplying by a test function and integrating over the domain (see section 4.4 of \cite{Melvin:2019}). For the pressure gradient term, integration by parts must be applied. Notably, the boundary-condition terms that arise from integration by parts can be ignored. This might seem counterintuitive given the importance of boundary conditions in a limited-area model. However, the philosophy here is to use the full global model equations while updating only within a subdomain. Consequently, at this point we only consider solving the equations over the globe, where the boundary condition terms are periodic and cancel each other out.

\subsection{Linear Solver}

When the finite element spatial discretisation is applied to Eq.~\ref{eq:iterated}, and reduced following the method and notation of \cite{Maynard:2020}, the following 2x2 block-matrix-form equation is obtained:
\begin{equation}
  \label{eq:mixed}
  \left(
  \begin{array}{ll}
     \overline{\bm M}_2^{\mu,C} & - \bm G^{\theta ^*} \\
     \overline{\bm D} & \bm M_3^{\Pi^*}
  \end{array}
  \right)
  \left(
  \begin{array}{l}
  \bm \xi_u \\
  \bm \xi_\Pi 
  \end{array}
  \right) =
  \left(
  \begin{array}{l}
  - \overline{\mathcal{R}}_u \\
    -\overline{\mathcal{R}}_{\Pi, \rho}
  \end{array}
  \right)
\end{equation}
where $\bm \xi_u$ and $\bm \xi_\Pi $ are the vectors of finite element coefficients for the increments $\mathbf{u}'$ and $\Pi'$. $\overline{\bm M}_2^{\mu,C}$ is the augmented $\mathbb{W}_2$ mass matrix (with damping layer, Coriolis terms and augmented with the buoyancy coupling between the vertical velocity and potential temperature), $\bm M_3^{\Pi^*}$ is the (block diagonal) $\mathbb{W}_3$ mass matrix scaled using $\Pi^*$, $\bm G^{\theta ^*}$ is the gradient operator (in weak form) scaled using $\theta^*$, which maps from $\mathbb{W}_3$ to $\mathbb{W}_2$, and $\overline{\bm D} $ is the augmented divergence operator which maps from $\mathbb{W}_2$ to $\mathbb{W}_3$.

Specifically,
\begin{equation}
  \begin{aligned}
  \overline{\mathcal{R}}_\Pi & = \mathcal{R}_\Pi - {\bm P}_{3\theta}\mathring{\bm M}^{-1}_\theta \mathcal{R}_\theta \\
  \overline{\mathcal{R}}_u & = \mathcal{R}_u - {\bm P}_{2\theta}\mathring{\bm M}^{-1}_\theta \mathcal{R}_\theta \\
  \overline{\mathcal{R}}_{\Pi, \rho} & = \overline{\mathcal{R}}_\Pi + \bm{M}_3^{\rho^*} \bm{M}_3^{-1}\mathcal{R}_\rho \\
  \end{aligned}
\end{equation}
and
\begin{equation}
  \begin{aligned}
\overline{\bm D}  & = \bm{M}_3^{\rho^*} \bm{M}_3^{-1} \bm{D}^{\rho^*}  + \bm P_{3\theta} \mathring{\bm{M}}_\theta^{-1} \bm{P}_{\theta 2} \\
\overline{\bm M}_2^{\mu,C} & = \bm{M}_2 + \tau_u \Delta t ( \bm{M}_\mu + \bm{M}_C ) + \bm P_{2\theta} \mathring{\bm{M}}_\theta^{-1} \bm{P}_{\theta 2} \\
  \end{aligned}
\end{equation}
where $\bm{M}_\theta$, $\bm{M}_3$ and $\bm{M}_2$ are the mass matrices associated with the $\mathbb{W}_\theta$, $\mathbb{W}_3$ and $\mathbb{W}_2$ spaces respectively, $\bm{M}_\mu$ and $\bm{M}_C$ are the mass matrices associated with the damping layer and Coriolis terms respectively. $\mathring{\bm{M}}_\theta^{-1}$ is the inverse of a (diagonal) mass-lumped version of $\bm{M}_\theta$.

The projection operators $\bm{P}_{\theta 2}$ and $\bm{P}_{2 \theta}$ only consider the vertical components. In addition, the coefficients associated with vertical-horizontal couplings are removed from $\overline{\bm M}_2^{\mu,C}$. It was found that the inclusion of these cofficients sometimes make the linear solver extremely slow to converge, especially in cases where there is steep orography. The removal of these terms has negligible effect on the final solution. This is because the couplings are still retained in the right-hand-side equations and so are still dealt with by the Newton iterations. Removing the terms from the left-hand-side just means that they are evaluated at the previous Newton iteration, rather than the current iteration.

The (2x2) system is solved using a Krylov method, with the associated approximate Helmholtz equation serving as the preconditioner. The approximate Helmholtz equation is derived by eliminating $\bm \xi_u$ and using a mass-lumped, diagonal approximation $\mathring{\bm M}_2$ for $\overline{\bm M}_2^{\mu,C}$, to give
\begin{equation}
  \label{eq:helmholtz}
  \begin{aligned}
  \bm H \bm \xi_\Pi & =  \bm B \\
 \text{where   } \bm H & = \bm M_3^{\Pi^*} +  \overline{\bm D} \mathring{\bm M}_2^{-1} \bm G^{\theta^*}  \\
 \text{and    } \bm B & = -\overline{\mathcal{R}}_{\Pi,\rho} +  \overline{\bm D} \mathring{\bm M}_2^{-1} \overline{\mathcal{R}}_u .
  \end{aligned}
\end{equation}

\subsection{Finite Volume Transport Discretisation}

A finite-volume scheme is used to discretise the transport terms $\mathcal{F}$, $\mathcal{A}^C$ and $\bm{\mathcal{A}}^A$ in Eq.~\ref{eq:rhs}. The one-dimensional Nirvana scheme \cite[]{Leonard:1995} is used with SWIFT (Splitting With Improved FFSL for Tracers) splitting, as described in \cite{Bendall:2024}, to give a three-dimensional Flux-Form Semi-Lagrangian (FFSL) scheme. Flux-form schemes compute fluxes at the cell boundaries to ensure conservation. FFSL schemes compute the flux over the timestep by integrating the field between the cell boundary and the departure point (i.e. the location where the flow originated).

Following the steps in \cite{Bendall:2024}, the advective form increment $\mathcal{A}^A(m, \mathbf{u})$ can be obtained by solving the conservative transport equation for $m$, and dividing by the conservative transport update of a unity field, similar to the approach of \cite{Putman:2007}. Similarly, the consistent form increment $\mathcal{A}^C(m,\rho,\mathbf{u})$ uses the $\rho$ fluxes to transport $m$, then divides by the updated $\rho$ to give the advective update of $m$. This ensures consistency between $m$ and $\rho$, and conservation of $m \rho$. 

Transport of vector fields $\bm{\mathcal{A}}^A$ is achieved by first transforming the vector field to orthogonal scalar components, applying the transport ${\mathcal{A}}^A$ to each scalar component, and then transforming back to the vector field (as described in \cite{Melvin:2024}).

The limiter of \cite{Zerroukat:2005} is applied to the Nirvana scheme to ensure monotonicity in one-dimension for potential temperature, with the SWIFT splitting enforcing this monotonicity for the 3D field. See \cite{Bendall:2024} for further details. 

\section{Dynamical core for a regional domain}

We now describe how the global dynamical core can be adapted to run on a regional domain. The global dynamical core can be summarised as solving 
 $\mathbf{x}_{k+1} =  \mathbf{x}_{k} + \mathbf{x}'$ iteratively, for $k=1,n$ (with $n=4$ used in this paper, giving 4 iterations). $\mathbf{x}$ is the state vector that contains the degrees-of-freedom (finite-element coefficients) for $\mathbf{u}$ and $\Pi$. On each iteration $k$, $\mathbf{x}'$ is obtained by solving the (2 x 2) linear system (equivalent to eq.~\ref{eq:mixed}):
\begin{equation} \label{eq:iteration}
  \mathcal{L}\left( \mathbf{x}^* \right) \mathbf{x}' = - \mathcal{R} \left( \mathbf{x}_{k} \right)
\end{equation}
where $\mathbf{x}^*$ is a reference state.

\begin{figure}[t!]
\begin{center}
\includegraphics[width=0.99\columnwidth]{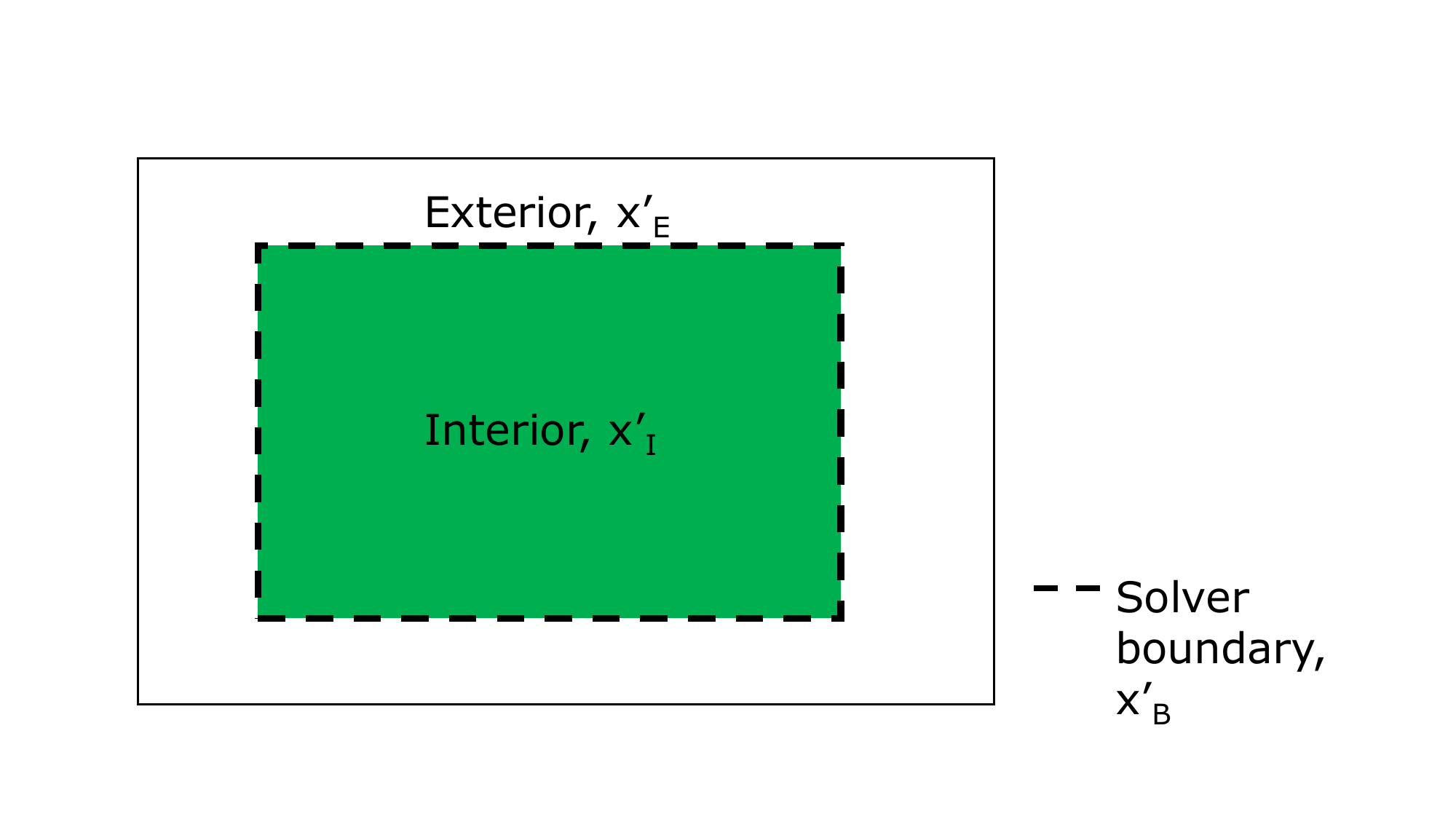}
\caption{{The domain is split into internal and external subdomains, divided by a boundary. 
{\label{fig:domain_ie}}%
}}
\end{center}
\end{figure}

Typically when a boundary-value-problem is solved using finite-elements, the inhomogeneous problem (with non-zero boundary conditions) is transformed into a homogeneous problem (with boundary conditions set to 0) by moving the boundary-terms to the right-hand-side and the trial functions are modified so that they vanish on the boundary e.g. \cite[]{Fix:1983,LarsonBengzon}.

The type of boundary condition also needs to be defined. The finite-element spaces in GungHo follow the de Rham complex \cite[]{Melvin:2019}, and these can be adjusted to encorporate homogeneous boundary conditions e.g. \cite[]{Boffi:2013}. In particular, the Raviart-Thomas space (Eq.~\ref{eq:fes}) becomes $\mathbf{v}_i \in \mathbb{W}_2^0$ which is the space of functions such that $\mathbf{v}_i \in \mathbb{W}_2$ and $\mathbf{v}_i \cdot \mathbf{n} =0 $ on the boundary. i.e. with Neumann boundary conditions. Thus, because of the definition of the mixed finite-element space formulation, the appropriate boundary condition is to specify the momentum increment $\mathbf{u}'$ on the boundary.

These approaches are also taken here, but indirectly rather than directly. To enable a more straightforward technical implementation, we emulate these approaches whilst using the framework of the full global model.

We wish to emulate the approach of transforming to a homogeneous problem. But we wish to do this whilst still using the full global mesh so that the right-hand-side terms (e.g. transport) can be calculated using the full mesh whilst the solver increment can be calculated only within the solver interior, without the need for transferring between meshes. This is reconciled by splitting the global solution vector  $\mathbf{x}'$ into the sum of a vector that has zeros everywhere other than the interior $\mathbf{x'}_I$, plus one that is zero everywhere other than the boundary $\mathbf{x'}_B$, plus one that is zero everywhere other than the exterior $\mathbf{x'}_E$:
\begin{equation}\label{eq:split}
\mathbf{x}' = \mathbf{x}'_I + \mathbf{x}'_B + \mathbf{x}'_E.
\end{equation}

These domains can be visualised in Fig.~\ref{fig:domain_ie}. The non-zero values of $\mathbf{x}'_B$ and $\mathbf{x}'_E$ are provided by the driving model. This splitting underpins the philosophy of only using the model to update the solution within the subdomain. Noting that Eq.~\ref{eq:iteration} is linear in $\mathbf{x}'$, it can be rearranged (using Eq.~\ref{eq:split}) as
\begin{equation}\label{eq:split_with_external}
  \mathcal{L}\left( \mathbf{x}^* \right) \mathbf{x}'_I = - \mathcal{R}\left( \mathbf{x}^{k} \right) - \mathcal{L}\left( \mathbf{x}^* \right) \mathbf{x}'_E - \mathcal{L} \left( \mathbf{x}^* \right) \mathbf{x}'_B .
\end{equation}
We emphasise that the splitting is only applied to the increments $\mathbf{x}'$ and is not applied to $\mathbf{x}^*$ or $\mathbf{x}^k$. The splitting is implemented through the use of a diagonal matrix ${\bm Z}$, which we refer to as a zeroing matrix. This multiplies the exterior and boundary values by zero, and the interior values by one so that only the interior values are retained:

\begin{equation}
  \mathbf{x}'_I = \bm{Z} \mathbf{x}'.
\end{equation}

We also emulate the approach of modifying the trial functions so that they vanish on the boundary. However, rather than modifying these within the actual operators, the standard (global) operators are used and then pre- and post-multiplication using the zeroing matrices gives an equivalent result. There are many operators that would need to be modified - and they would only need to be modified for use only in the solver and not in the transport (which uses the full mesh). Thus the use of the zeroing matrices results in less code, making it easier to implement and maintain.

\subsection{Implementation details}

\begin{figure}[t!] 
\includegraphics[trim={0 6cm 0 0},clip,width=0.99\columnwidth]{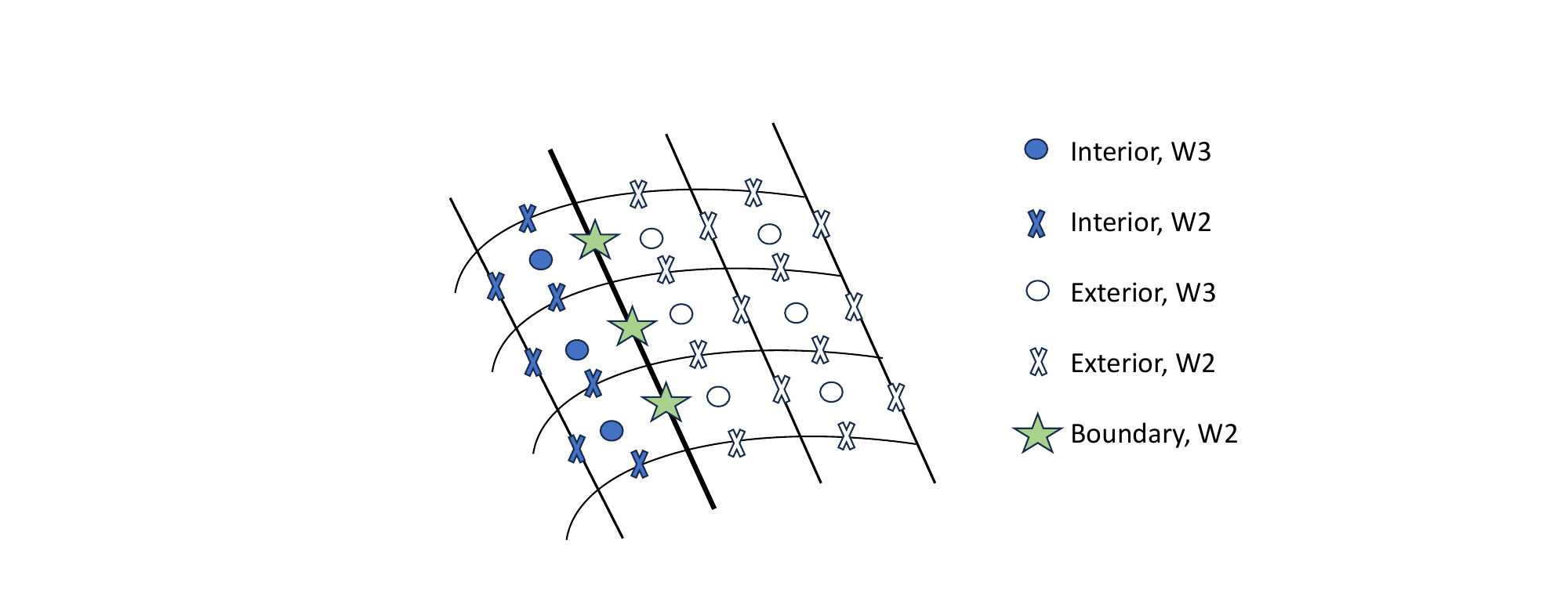}
\caption{The location of the degrees-of-freedom (dofs) for finite-elements in $\mathbb{W}_2$ and $\mathbb{W}_3$, and their relation to whether they are in the interior, exterior or on the boundary.}
\label{fig:zero}
\end{figure}

The splitting is created by subsetting the finite-element basis-functions so that there are $N^\Pi_I +  N^\Pi_E + N^\Pi_B$ basis functions for $\Pi$ and $N^u_I +  N^u_E + N^u_B$ basis functions for $\mathbf{u}$ in total, and
\begin{equation}
  \begin{aligned}
  \mathbf{u}'_B(x) & = \sum_{i=1}^{N^u_B} \xi_u^i \mathbf{v}_i (x) &
  \Pi'_B(x) & = \sum_{i=1}^{N^\Pi_B}  \xi_\Pi^i \sigma_i(x) \\
  \mathbf{u}'_E(x) & = \sum_{i=N^u_B + 1}^{N^u_E} \xi_u^i \mathbf{v}_i (x) &
  \Pi'_E(x) & = \sum_{i=N^\Pi_B + 1}^{N^\Pi_E}  \xi_\Pi^i \sigma_i(x) \\
  \mathbf{u}'_I(x) & = \sum_{i=N^u_E +1}^{N^u_I} \xi_u^i \mathbf{v}_i (x) &
  \Pi'_I(x) & = \sum_{i=N^\Pi_E + 1}^{N^\Pi_I}  \xi_\Pi^i \sigma_i(x). \\
  \end{aligned}
\end{equation}

Thus the split vectors, with $\bm \xi_u = \bm \xi_u^B + \bm \xi_u^E + \bm \xi_u^I$ are now:

\begin{equation}
  \begin{array}{lllllllllll}
  \bm \xi_u^B  = [ & \xi_u^1 & \dots & \xi_u^{N_B^u} & 0 & \dots & 0 & 0 & \dots & 0 & ]^T \\
  \bm \xi_u^E  = [ & 0 & \dots & 0 &  \xi_u^{N_B^u+1} & \dots & \xi_u^{N_E^u} & 0 & \dots & 0 & ]^T \\
  \bm \xi_u^I  = [ & 0 & \dots & 0 &  0 & \dots & 0 & \xi_u^{N_E^u+1} & \dots & \xi_u^{N_I^u} & ]^T
  \end{array}
\end{equation}
and similarly for  $\bm \xi_\Pi$ and so that
\begin{equation}
  \mathbf{x}'_I = \left[ \begin{array}{l}
      \bm \xi_u^I \\
      \bm \xi_\Pi^I
      \end{array}
      \right],
\end{equation}
and similarly for $\mathbf{x}'_E$ and $\mathbf{x}'_B$.

We define $\mathbf{u}'_B(x)$ using the values of $\xi_{u}$ associated with the solver boundary. There are no basis functions associated with the boundary contributing to the pressure increment, so $N^\Pi_B=0$, and $\Pi'_B(x)=0$. The location of the degrees-of-freedom associated with this splitting is illustrated in Fig.~\ref{fig:zero}.

\begin{figure}[t!] 
\includegraphics[scale=0.6]{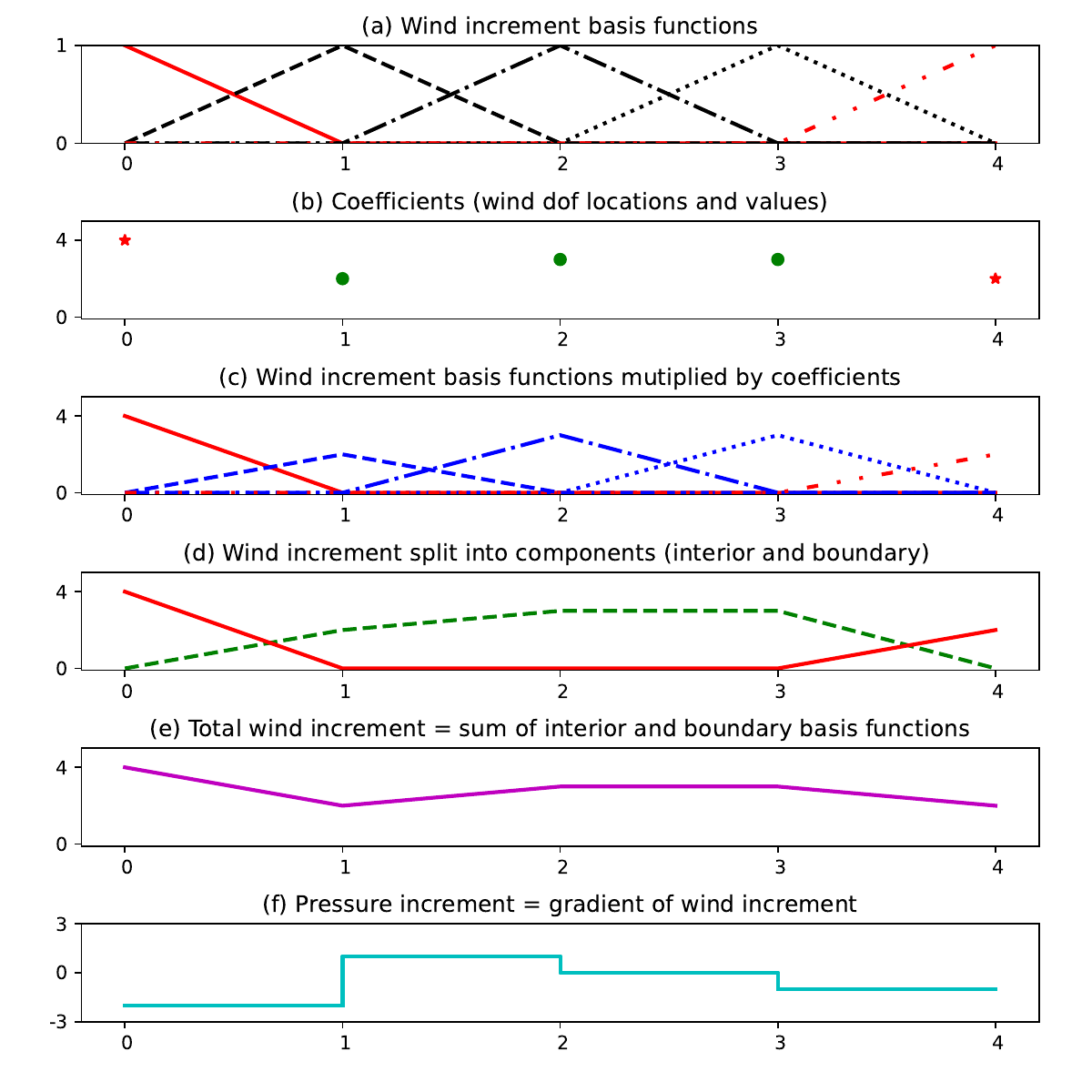}
\caption{Diagram of example 1D basis functions and solutions from the mixed system. (a) The basis functions for $\mathbb{W}_2$ (b) example coefficients (c) the basis functions multiplied by the coefficients  (d) splitting into the interior $u'_I \in \mathbb{W}_2$ (green dash) and boundary $u'_B \in \mathbb{W}_2$ components (red solid) (e) the total solution $u' = u'_I + u'_B$, and (f) the total solution for pressure $\Pi' \in \mathbb{W}_3$. The boundaries are located at $x=0$ and $x=4$.}
\label{fig:basis}
\end{figure}

The use of this splitting, together with the basis-functions, is illustrated in Fig.~\ref{fig:basis}, which considers a simple 1D case where the right-hand-side forcing $\mathcal{R}\left(\mathbf{x}^k \right)$ is zero. When the boundary conditions are prescribed by the basis functions associated with $\mathbf{u}'$ at the boundary, they influence both the interior pressure and wind (as the basis functions share the same local support). The basis functions associated with the exterior are ignored as they do not influence the interior.

Equation~\ref{eq:split_with_external} also contains a term corresponding to the external region, and this has an influence on the boundary term (which is fixed). But, with the above definition of splitting, the external term can be removed by pre-multiplying both sides by $\bm{Z}$ so that $\bm{ Z} \mathcal{L} \mathbf{x}_E=0$, and Eq.~\ref{eq:split_with_external} reduces to:
\begin{equation}
  \bm{Z} \mathcal{L}  \mathbf{x}'_I = - \bm{ Z} \mathcal{R}\left( \mathbf{x}^k \right)  - \bm{Z} \mathcal{L} \mathbf{x}'_B \equiv \mathbf{y}
\end{equation}

This is now a singular system meaning that $\tilde{\mathcal{L}} = \bm{Z} \mathcal{L} \bm{Z}$ is not invertible. However, this is solved using a Krylov method (Generalized Conjugate Residual), with the equivalent least-squares problem 
\begin{equation}
  \min_{\mathbf{x}' \in \mathbb{R}} \| \bm{Z} \mathcal{L} \bm{Z} \mathbf{x}' - \mathbf{y} \|_2.
\end{equation}
This is guaranteed to converge as the null space of $\tilde{\mathcal{L}}$ is the same as the null space of $\tilde{\mathcal{L}^T} $ \cite[]{Hayami:2011}. In effect we are solving for the non-zero parts of $\mathbf{x}'_I$, and the parts set to zero (i.e. corresponding to the boundary and exterior) remain zero. It could also be possible to create a smaller system, with a smaller vector containing only the non-zero parts of $\mathbf{x}'_I$, but this is not necessary and it is easier technically to retain the zero part and solve using the full-sized vector. It would also be possible to modify the terms in $\mathcal{L}$ to give a new operator $\tilde{\mathcal{L}} =  \bm{Z} \mathcal{L} \bm{Z}$ but they are kept separate here for ease of technical implementation.

The zeroing matrix can be described as two separate matrices: $\bm Z_2$ for the variables in $\mathbb{W}_2$ and $\bm Z_3$ for the variables in $\mathbb{W}_3$, so that
\begin{equation}
  \bm Z= \left( \begin{array}{cc}
    \bm Z_2 & \mathbf{0} \\
    \mathbf{0} & \bm Z_3 \\
    \end{array} \right)
\end{equation}
where $\mathbf{0}$ is a matrix containing only zeros.

For example,
\begin{equation} \label{eq:mask}
  \bm Z_3 = diag(\bm b)
\end{equation}
is diagonal with the elements of the vector $\bm b = [1, 1, 1, ... , 0, 0, 0, ...]$ on the diagonal. We refer to ${\bm Z}_3$ as the $\mathbb{W}_3$ zeroing matrix, and $\bm b$ as a binary mask. An example of a binary mask for the $\mathbb{W}_3$ space is shown later on in the results section, in Fig.~\ref{fig:blend_weights}(a). Further motivation for this approach and the technical details on how the binary masks are specified are provided in the appendix. 

 The zeroing-matrix $\bm Z_2$ is used to multiply the variables in $\mathbb{W}_2$ in the exterior and on the boundary by zero, and $\bm Z_3$ is used to multiply the variables in $\mathbb{W}_3$ in the exterior by zero. These can be applied to the full 2x2 matrix (\ref{eq:mixed}), to give:
\begin{equation}\label{eq:mixed_lam}
\begin{aligned}
\left[ \begin{array}{cc}
 \bm Z_2 \overline{\bm M}_2^{\mu,C} & - \bm Z_2 \bm G^{\theta^*}  \\
 \bm Z_3 \overline{\bm D} & \bm Z_3 \bm M_3^{\Pi^*}
\end{array} \right]
\left[\begin{array}{c}
\bm {\xi}_u^I \\ \bm {\xi}_\Pi^I \\ \end{array} \right] 
& =\left[\begin{array}{c}
\overline{\bm r}_u \\  \overline{\bm r}_\Pi \\ \end{array} \right]
\end{aligned}
\end{equation}
where $\overline{\bm r}_u = \bm Z_2 ( \overline{\mathcal{R}}_u - \overline{\bm M}_2^{\mu,C} \bm \xi^B_u)$ and $ \overline{\bm r}_\Pi = \bm Z_3 (\overline{\mathcal{R}}_{\Pi,\rho} -  \overline{\bm D} \bm \xi^B_u)$ and $\bm \xi^B_u$ are the known values of the finite-element coefficents associated with $\mathbf{u}'$ on the boundary of the solver (from the driving model LBCs). This is the same as the linear solver for the global domain, except that the right hand side has been augmented by the boundary term, and the zeroing matrices are applied to both sides. We can see that the boundary conditions $\bm \xi_B^u$ provide a new right-hand-side for both 
the $u$-equation (through the $\mathbb{W}_2$ mass matrix $\overline{\bm M}_2^{\mu,C}$), and the $\Pi$-equation (through the divergence $\overline{\bm D}$ ).

Using this new right-hand-side, the same elimination procedure as that without boundary conditions can be used to form the Helmholtz equation (Eq.~\ref{eq:helmholtz}), but with the additional application of the $\bm{Z}$ matrices.
\begin{equation}\label{eq:HelmholtzLAM}
 \bm H = \bm Z_3 \left( \bm M_3^{\Pi^*} +  \overline{\bm D} \bm Z_2 \mathring{\bm M}_2^{-1} \bm Z_2 \bm G^{\theta^*} \right) \bm Z_3
\end{equation}

The $\bm{Z}$ matrices are also applied in the associated back substitution for the momentum increment
\begin{equation}
  \bm {\xi}_u = \bm Z_2 \mathring{\bm M}_2^{-1}( \bm G^{\theta^*} \bm {\xi}_\Pi - \bm r_u).
\end{equation}

We can see here that as $\mathring{\bm M}_2^{-1}$ is diagonal, this Helmholtz equation is the same equation as would be obtained if Neumann boundary conditions (gradient of $\Pi$ normal to the boundary) were added directly to the Helmholtz equation.

In summary, the use of the iterated-semi-implicit timestepping results in the requirement to solve a linear system. This linear system requires boundary conditions when solving over a regional domain. These boundary conditions have been applied by augmenting the right-hand-side with terms that include the wind increment on the boundary. Then, using zeroing matrices to effectively modify the operators, only the interior solution is solved for whilst using the same framework (mesh and operators) as the global model.

\begin{figure}[t!]
\begin{center}
\includegraphics[width=0.99\columnwidth]{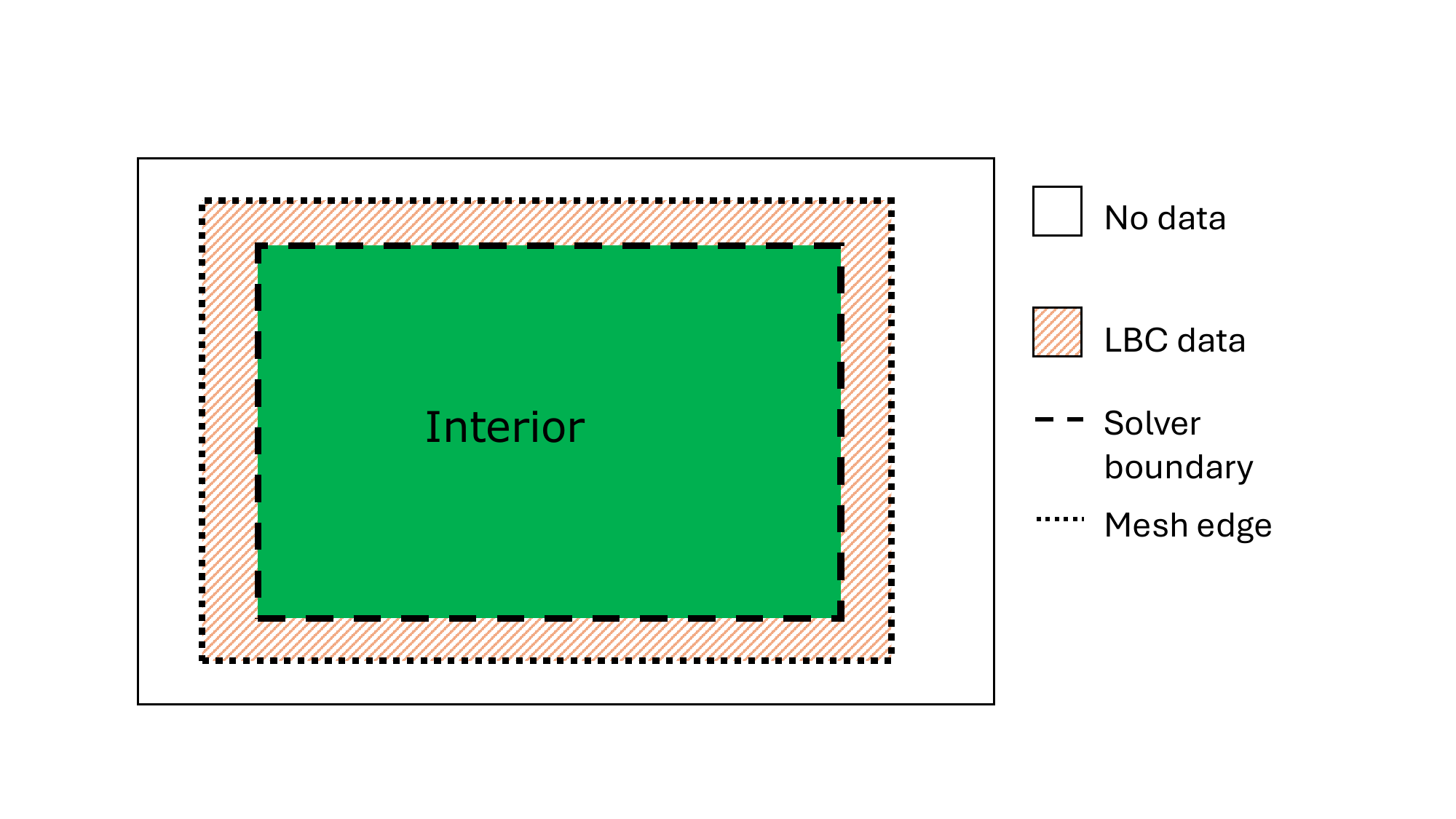}
\caption{{LBC data is supplied in some of the external region. The solver boundary is the same as that in Fig.~\ref{fig:domain_ie}. 
{\label{fig:domain_rim}}%
}}
\end{center}
\end{figure}

\subsection{LBC data}
The right-hand-side (RHS) terms $\mathcal{R} \left( \mathbf{x}^{k} \right)$ of Eq.~\ref{eq:iteration} include transport. In the physical problem, when the wind advects from the exterior to the interior, information is required on the upwind side of the internal subdomain - i.e. information from the external subdomain. Thus, whilst the linear solver is used to update the increments in the interior, another approach is required to update the values in the exterior, such as overwriting with lateral boundary condition (LBC) data from a driving model. The computation of the RHS terms (including the transport scheme) is then exactly the same as the full global model.

The LBC data does not need to cover the whole exterior; it only needs to be specified in a small region around the interior, as illustrated in Fig.\ref{fig:domain_rim}. The minimum depth of this region will depend on the particular transport scheme, the cell-size and the timestep. For computational memory efficiency, we require an LBC region of the smallest depth possible. This depth should be chosen so that the advective tendencies in the interior subdomain (and on the boundary) can be specified without requiring data beyond the LBC data region.

\subsection{Blending}
Ideally, the LBC data updated in the exterior should come from a case where the numerical model has been run over the entire domain at the same resolution and with the same configuration as the regional model. But of course, this would negate the point of having a regional model! The purpose of having a regional model is to refine the resolution over the region of interest, so that the LBCs are provided by a lower-resolution model.

This difference between the lower-resolution driving model and the higher-resolution regional model leads to inconsistencies between the values in the interior and the LBCs.  However, blending can be applied to minimize these mismatches. In the literature, there are two main methods to achieve this: in Rayleigh damping or nudging \cite[]{Davies:1976, Skamarock:2018, Pham:2021} the LBC solution is added directly to the model prognostic equations as an additional forcing term with a relaxation coefficient; in blending \cite[]{Davies:2014}, the model solution is blended with the LBC solution at the end of the timestep (or substeps). This latter blending method is really a discretisation of the nudging method. Both methods give a similar result, and essentially only differ in how they are implemented. Following the success of using blending in the Met Office Unified Model and the ease of technical implementation,  we choose to use the \cite{Davies:2014} approach of blending.

\begin{figure}[t!]
\begin{center}
\includegraphics[width=0.99\columnwidth]{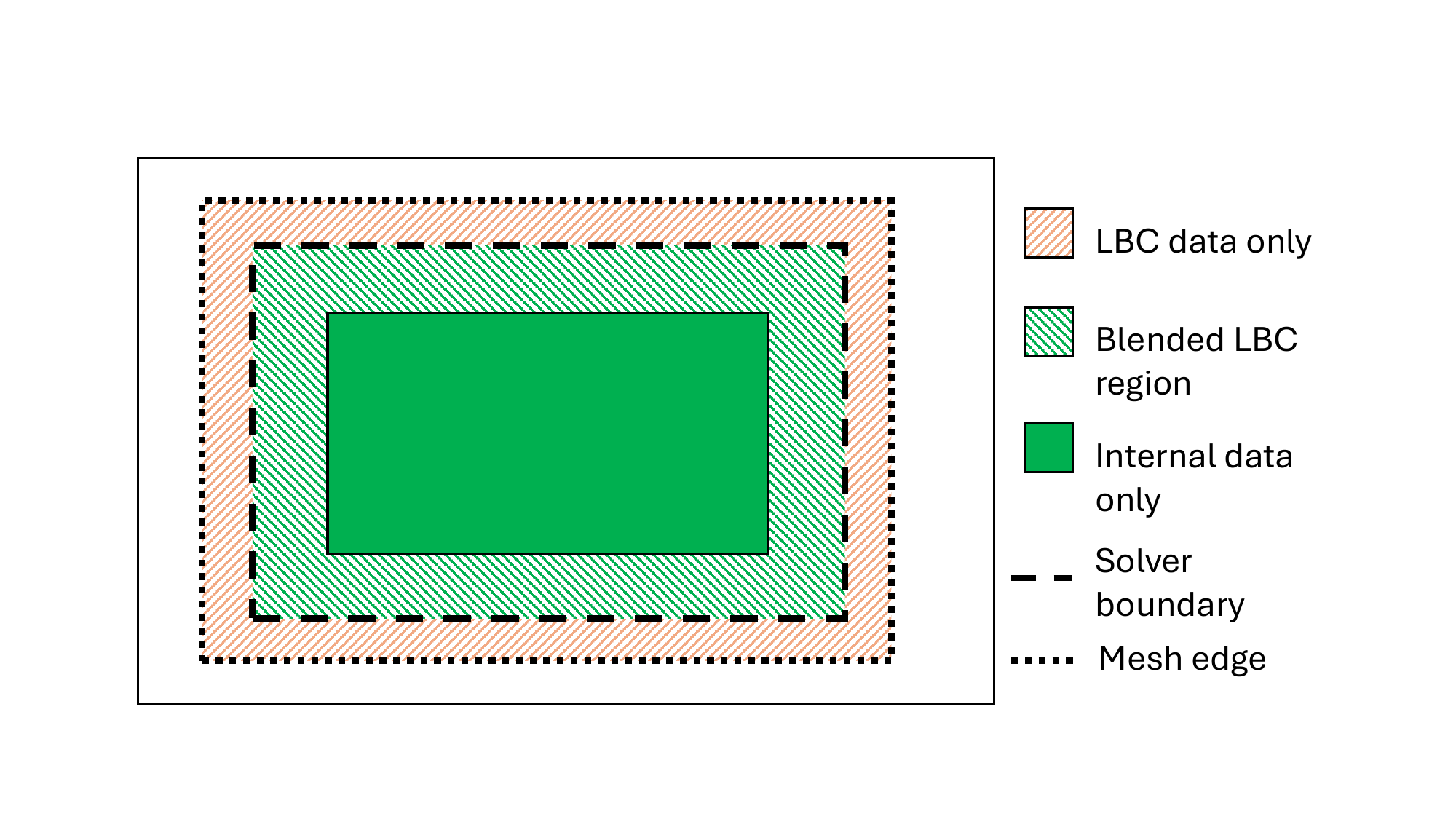}
\caption{{LBC data is supplied in some of the internal region, and is blended with the interior solution.
{\label{fig:domain_blend}}%
}}
\end{center}
\end{figure}

In this approach, LBC data is also provided in a few cells in the interior of the domain and this is blended with the interior solution, as shown in Fig.\ref{fig:domain_blend}. The blending procedure is a simple linear interpolation between the LBC data and the interior solution, using specified blending weights:
\begin{equation}
   x_{new} = (1-w) x_{current} + w x_{LBC}.
\end{equation}
The blending weights ramp down from one to zero, from the solver boundary and towards the middle of the domain. Different blending weights are used for variables in $\mathbb{W}_2$ to the variables in $\mathbb{W}_3$: the weights for a cell face in $\mathbb{W}_2$ are an average of the $\mathbb{W}_3$ weights specified in the adjoining cells.

The overwriting with the LBC data and the blending in the blend region is applied after every iterated-semi-implicit iteration to every prognostic variable ($\Pi, \mathbf{u}, \rho, \theta)$. This ensures that the interior solution is always consistent with the LBCs.

There is no theory as to what the optimal blending weights should be as it is really an ad-hoc approach to manage the incompatibility of the boundary conditions. However, the best values will likely depend on the spatial and temporal resolution of the LBC data and of the regional model, and in practice are found through numerical testing.

\subsection{Multigrid Solver}
To solve the mixed system of equations (Eq.~\ref{eq:mixed_lam}), an iterative Krylov based method is used; in practice this is a generalised-conjugate-residual (GCR) scheme. This uses an approximate Schur-complement preconditioner that involves solving the Helmholtz equation Eq.~\ref{eq:HelmholtzLAM} at each step. As the Helmholtz equation is only used as a preconditioner for the mixed system, only an approximate solution is required. Therefore, although it is possible to use an iterative method to solve the Helmholtz equation, in practice it is found to be sufficient to only use a single V-cycle of a multigrid-method, as described in \cite{Maynard:2020}.

In the multigrid algorithm, the general idea is that the solution is solved on progressively coarser meshes and then back to the fine mesh, with some cycling back and forth between the coarse and fine meshes. As defined by \cite{Maynard:2020}, there are 2 key operators to map between the meshes: prolongation and restriction. To enable multigrid to be used by the regional model, these operators can be adapted through the use of the binary masks, $\bm b$ (Eq.~\ref{eq:mask}).

To transfer from the coarse mesh $l$ to the fine mesh $l+1$, the prolongation operator is used. For a given coarse cell $C$, define the corresponding fine cells f = 1, 4 (that fit within the coarse cell)
\begin{equation}
  {\Pi'_f}^{l+1} = b^{l+1}_f {\Pi'}_C^{l}
\end{equation}
where the binary mask on the fine mesh $b^{l+1}_f$ is $0$ or $1$,
so that a cell in the fine mesh is given by the value in the corresponding coarse mesh and the cells corresponding to a binary mask value of zero are assigned a value of zero.

To transfer from the fine mesh to the coarse mesh, the restriction operator is used. For given fine cells f = 1, 4, define the corresponding coarse cell $C$ as
\begin{equation}\label{eq:restrict}
  r^l_C = \left\{
  \begin{array}{ll}
  \frac{\sum_f b_f^{l+1} r_f^{l+1}}{ \sum_f b_f^{l+1}} & \text{for   }  \sum_f b_f^{l+1} >0 \\
  0 &  \text{for   } \sum_f b_f^{l+1} = 0
  \end{array} \right.
\end{equation}
so that a cell in the coarse mesh is given by an average of the corresponding non-zero cells in the fine mesh. Cells corresponding to a binary mask value of zero are not included in the average so that missing-data is not included. When the binary mask field is entirely ones, these both revert back to the standard operators.

\begin{figure}[t!]
\begin{center}
\includegraphics[width=0.99\columnwidth]{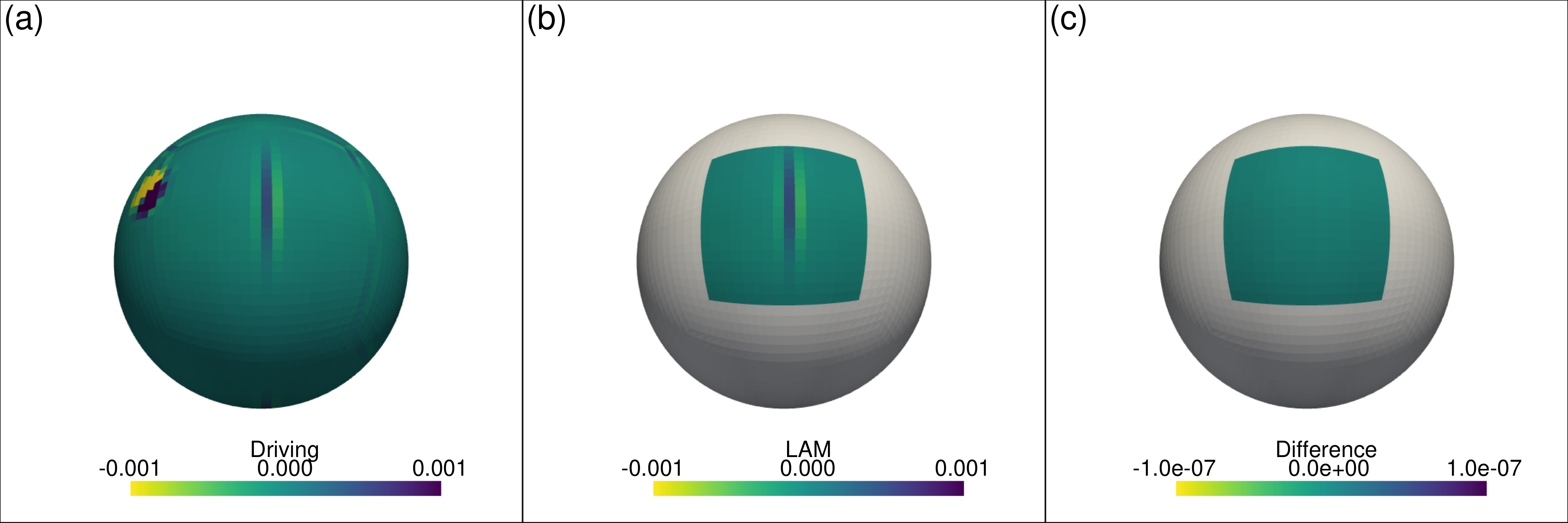}
\caption{Surface potential temperature $(^{\circ}C)$ with initial conditions subtracted, shown after a single solve, for the a) driving model, b) regional model and c) driving model minus regional model. The binary mask that is used to define the regional model is coloured white. The mesh is shown by the black lines.
{\label{fig:onetimestep}}%
}
\end{center}
\end{figure}

With these modified definitions of the prolongation and restriction operators, we now need the binary mask fields to be defined for each multigrid level. These are defined using the restriction operation, replacing $r_f$ in Eq.~\ref{eq:restrict} with $b_f$, to give
\begin{equation}
  b^l_c = \left\{
  \begin{array}{ll}
  \frac{\sum_f b_f^{l+1} }{ \sum_f b_f^{l+1}} & \text{for   }  \sum_f b_f^{l+1} >0 \\
  0 &  \text{for   } \sum_f b_f^{l+1} = 0
  \end{array} \right.
\end{equation}
which takes the value 1 if any of the fine mesh cells are 1, and 0 otherwise. Whilst the multigrid method is only applied to a $\mathbb{W}_3$ field (i.e. to solve for pressure), binary masks for the $\mathbb{W}_2$ space are also required so that the Helmholtz operator (required by multigrid) can be defined on the coarse meshes. These are defined in a similar way to those for $\mathbb{W}_3$, through the use of the restriction operation. However, as degrees-of-freedom (dofs) are shared between cells for $\mathbb{W}_2$ the averaging is applied over the 2 adjoining coarse cells associated with a particular dof.

The use of the binary masks is a very simple technique that can be used to enable the multigrid method to be used by the regional model and means that missing data is not used in the calculations. There is no restriction on the location of the solver boundary in relation to the coarsest level mesh; the method will still work if the solver boundary bisects a coarse-level cell. In this case, the position of the boundary between the 1s and 0s in the binary mask fields actually moves outwards, away from the interior, as the mesh is coarsened.

\section{Idealised experiments}
The aim of these idealised experiments is to verify that the regional numerical model is working correctly. All experiments are setup as a big-brother type experiment, as described in \cite{Davies:2014}, where the driving model provides LBCs for the regional model at every timestep. The use of idealised tests (rather than full weather model tests) provides the ideal situation where LBCs can easily be provided at every timestep, without the need for time interpolation, and with identical models. Similar idealised experiments for nested models have also been studied by \cite{Park:2014}.

Ultimately the operational regional model will use a rotated-pole, lat-lon mesh, with the LBCs being regridded from the cubed-sphere mesh used by the driving-model. But in the following idealised studies the regional model is run on exactly the same mesh as the driving model, with the boundary conditions being imposed to create a subdomain using the zeroing-matrices and adapted solver. This approach eliminates the need to regrid the LBCs from the driving model mesh to the regional model mesh, as they share the same mesh. Furthermore, since the regional model uses the same spatial and temporal resolution and the same model configuration, its solution should be identical to that of the driving-model.

\subsection{Baroclinic wave tests}
The first experiment is based on the deep baroclinic wave test devised by \cite{Ullrich:2014}, using an equiangular cubed-sphere mesh \cite[]{Nair:2005} and similar to that shown in \cite{Melvin:2024} but with a rotated C24L30 mesh (24 x 24 cells on each panel, and 30 vertical levels) and a 1 hour timestep. The mesh is rotated so that the centre of a panel is at longitude 0$^{\circ}$, latitude 45$^{\circ}$. It uses lowest-order finite-elements, and a spherical coordinate system (the same coordinate system that is used to define the cubed-sphere). The initial wind perturbation is centered at longitude 290$^{\circ}$, latitude 40$^{\circ}$.

The regional version is run on a single panel of the cubed-sphere, where the centre of the panel is at longitude 0$^{\circ}$, latitude 45$^{\circ}$. This position means that the initial perturbation is not seen by the regional model; the perturbation must be propagated into the regional model through the LBCs as the model evolves. The solver boundary is applied 4 cells from the edge of the panel, so that there are 24-2*4=16 cells along the edge of the regional model domain. The cells on the outer edge of the solver boundary (i.e. 4 cells depth) are supplied with the LBC data. A depth of 4 cells is sufficient to enable the FFSL transport scheme to update the cells in the interior accurately for this spatial resolution and timestep.

\begin{figure}[t!]
\begin{center}
\includegraphics[width=0.99\columnwidth]{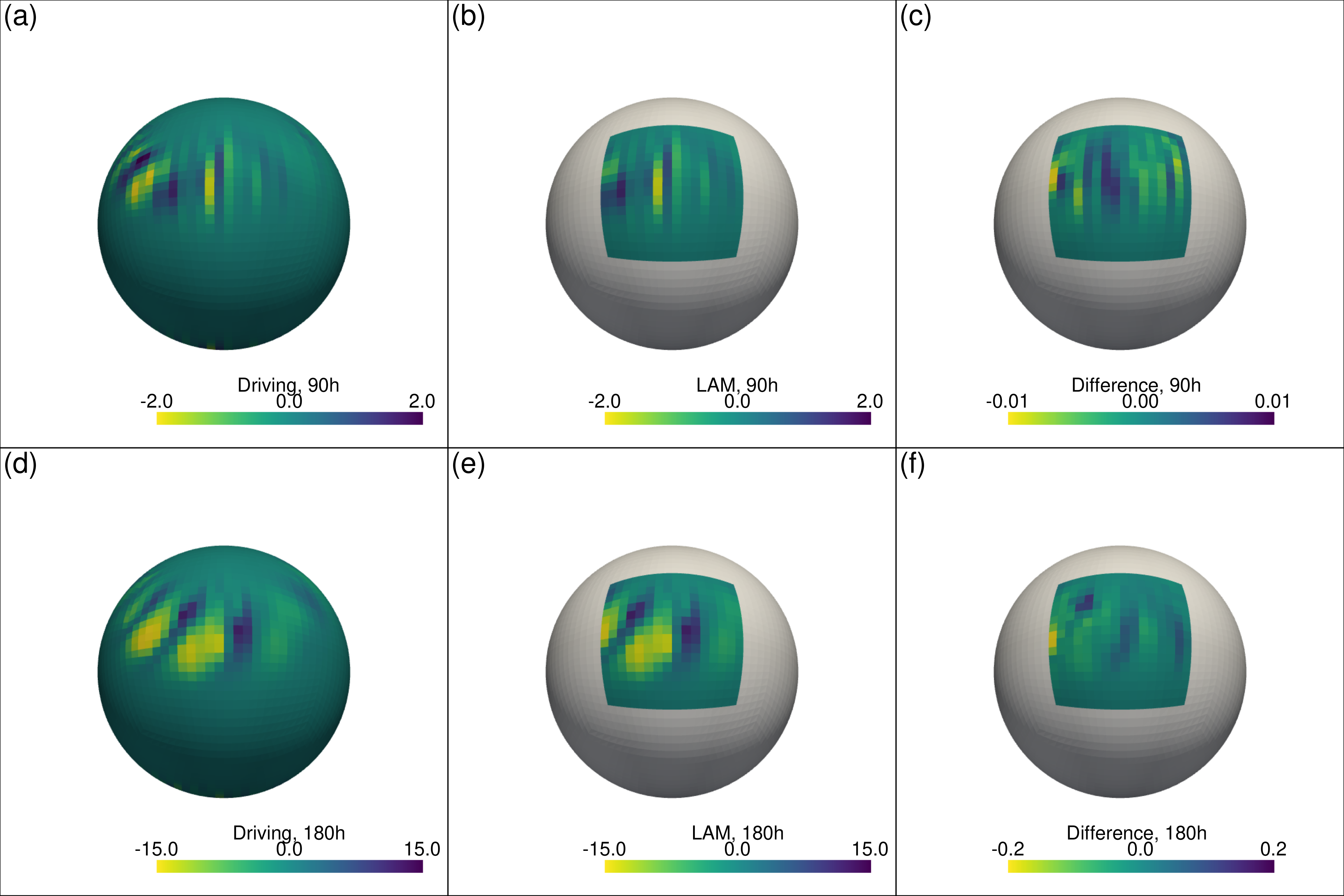}
\caption{Evolution of the surface potential temperature $(^{\circ}C)$ with initial conditions subtracted. (a-c) 90 hours and (d-f) 180 hours. Left: driving, Centre: regional, Right: regional minus driving.
{\label{fig:baroclinic}}%
}
\end{center}
\end{figure}

\subsubsection{Baroclinic wave: Single Timestep}
In the first experiment, both the driving model and regional model are run for a single timestep, with only 1 Newton iteration (n=1 in Eq.~\ref{eq:iteration}, rather than the usual n=4). There is no blending of the LBC data after the timestep so that the output at the end of the timestep is purely from the solver. In this situation the initial conditions and LBCs are exactly the same as the driving model. Therefore we would expect that the regional model should give exactly the same solution as the driving model, to within the solver tolerance. For both the driving and regional models, the solver for the mixed-system uses a relative tolerance of 10$^{-6}$ to ensure that the solver has fully converged. For both the driving and regional models, 64-bit precision (16 decimal digits) is used for all parts of the model, except for the solver which uses 32-bit (8 decimal digits) precision.

The results are shown in Fig.~\ref{fig:onetimestep}. By eye, the results from the driving and regional models look exactly the same, and the difference plot has a maximum amplitude less than 10$^{-7}$ which is consistent with the solver tolerance. Thus the regional model solution is the same as the driving model solution, to within the solver tolerance.

\subsubsection{Baroclinic wave: 7 day forecast}
In the second experiment, the models are run with a 1-hour timestep for 180 timesteps (7.5 days), with the data outside the solver boundary being overwritten by the LBC data at every timestep. The LBC data is only available at every timestep, rather than every Newton iteration - this gives a slight discrepancy between the driving model and regional model on each solve. $\mathbf{u}'$ on the boundary is no longer the same as the equivalent value in the driving model at that stage. It is instead specified as the difference betwen the value of $\mathbf{u}$ at the end of the driving model timestep, and the current value of $\mathbf{u}$ in the regional model. The solver tolerance for the mixed-system is specified as 10$^{-3}$ which is the default value for GungHo and has been found to be sufficient for accurate predictions without compromising on computational efficiency. Blending is not required as the LBCs are applied at every timestep, and the driving model is the same as the regional model.

 The results are shown in Fig.~\ref{fig:baroclinic}. In the driving model, the perturbation propagates from West to East, until a distinctive baroclinic wave appears. As expected, this is replicated in the regional model, with the information being passed through the LBCs into the interior. By eye, the regional model solution looks the same as that of the driving model. But the difference plots show that the differences between the regional and driving are now much larger than the numerical precision and there is some visible structure. These small differences are expected due to the increased solver tolerance for the mixed-system, the LBC data only being supplied every timestep (rather than every iteration), and the growth of errors over time.

\begin{figure}[t!]
\begin{center}
\includegraphics[width=0.7\columnwidth]{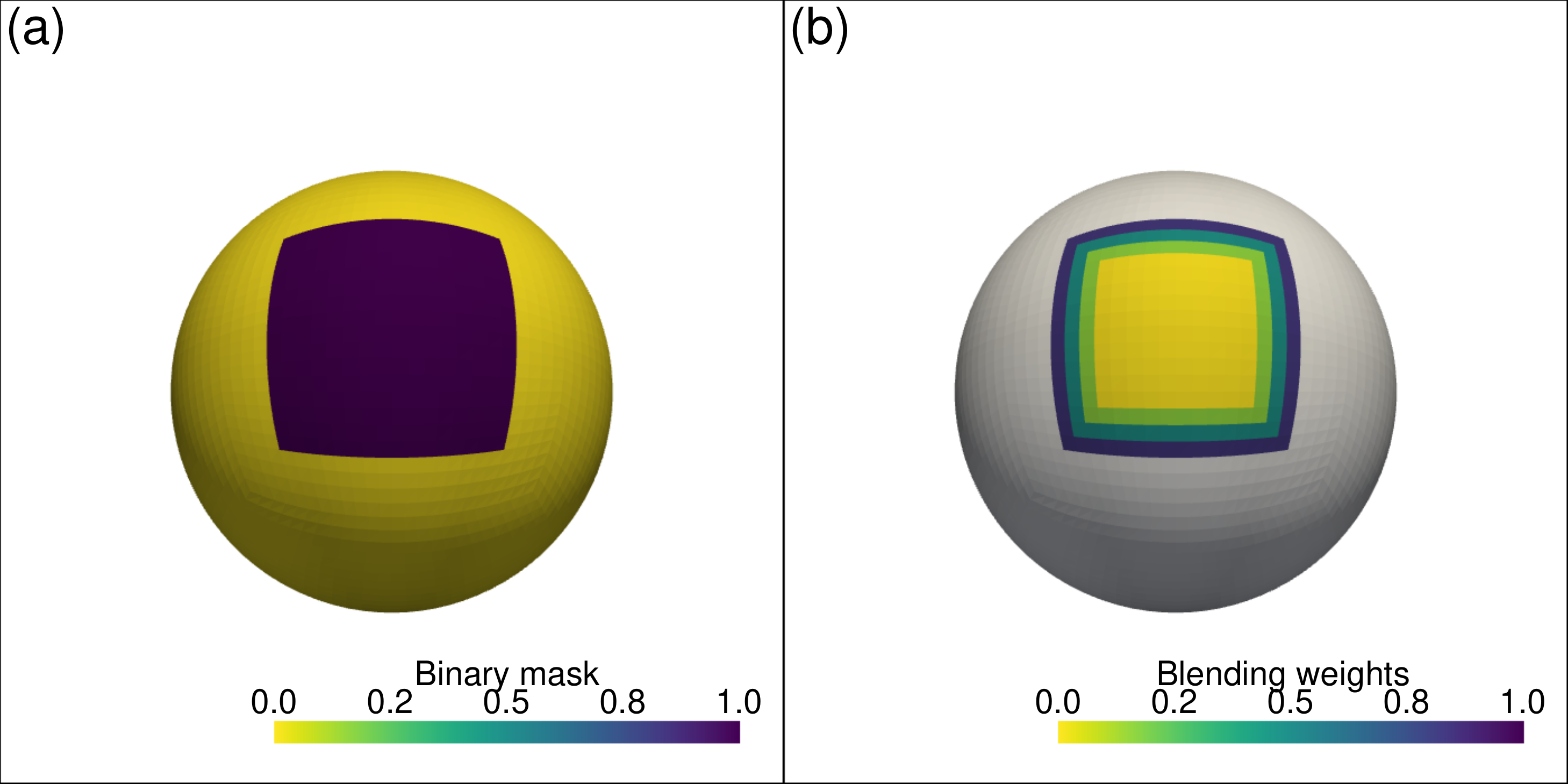}
\caption{{a) Binary mask that defines the position of the LAM domain and lateral boundaries and b) blending weights. Both can be used for the $\mathbb{W}_3$ and $\mathbb{W}_\theta$ fields
{\label{fig:blend_weights}}%
}}
\end{center}
\end{figure}

\subsubsection{Baroclinic wave: Time-interpolated LBCs and blending}
In the third experiment, the models are run with a 1-hour timestep for 300 timesteps, but the LBC data is no longer supplied every timestep. LBC data is instead supplied from the driving model every 60 timesteps and a linear time interpolation is used for the intermediate timesteps. This setup is more similar to a real weather prediction scenario, and it leads to an inconsistency between the regional model solution and LBC data that grows over the forecast period.

The inconsistency between the regional model and LBCs in this experiment allows us to test the impact of adding blending. An example of the blending weights used here is shown in Fig.~\ref{fig:blend_weights}; they are 3 concentric rings with weights that vary linearly from 1 to 0. There is no theory regarding what the optimal weights should be, so other weights could be chosen. But the blending weights specified here are sufficient for us to test the overall impact of applying blending.

\begin{figure}[t!]
\begin{center}
\includegraphics[width=0.99\columnwidth]{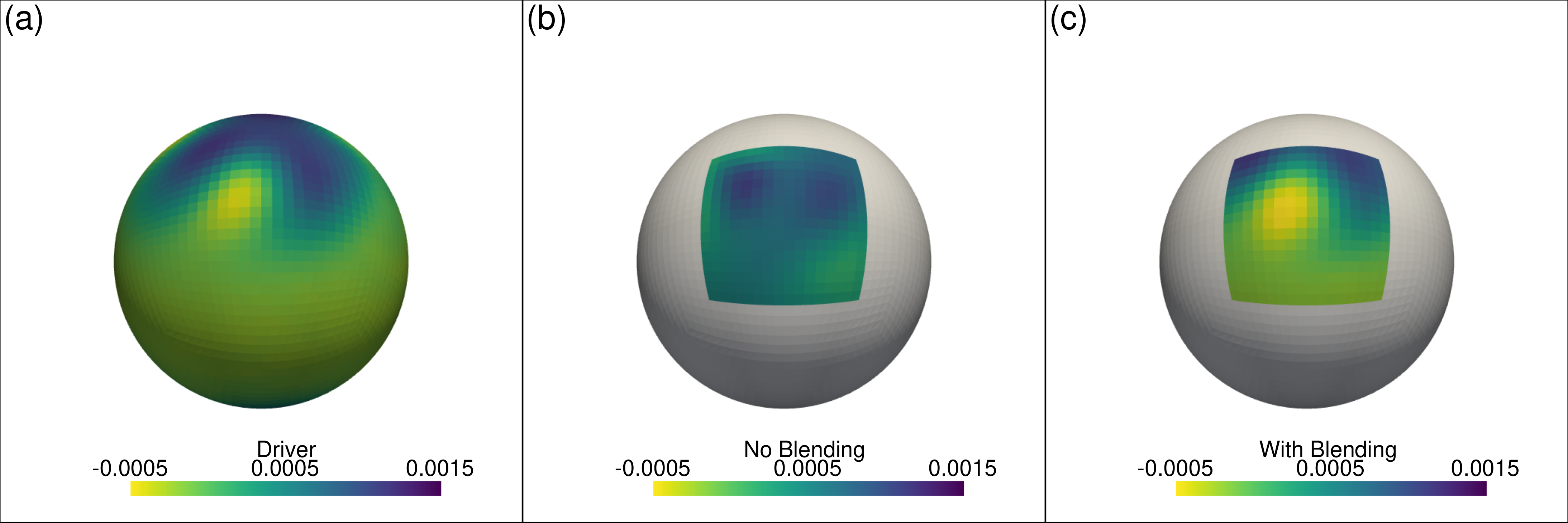}
\caption{{Exner pressure (Pa) with initial conditions subtracted, at 25km height and after 300 hours (12 days) simulation and the regional model using time-interpolated LBCs for the a) driving model b) regional model with no blending and c) regional model with blending.
{\label{fig:blend_pressure}}%
}}
\end{center}
\end{figure}

In this idealised experiment, the impact of adding the blending weights is most noticeable in the pressure field near to the top of the model, as shown in in Fig.~\ref{fig:blend_pressure}. Without blending (Fig.~\ref{fig:blend_pressure}b), a large discontinuity has developed at the boundary between the interior and the LBC region. In addition, the solution in the interior no longer resembles the solution of the driving model (Fig.~\ref{fig:blend_pressure}a). With blending (Fig.~\ref{fig:blend_pressure}c), there is no longer a discontinuity between the boundary and the LBC region and the solution in the interior looks more similar to the driving model. This gives a high level of confidence that the addition of blending helps to control the impacts of inconsistencies at the lateral boundaries.

\begin{figure}[t!]
\begin{center}
\includegraphics[width=0.99\columnwidth]{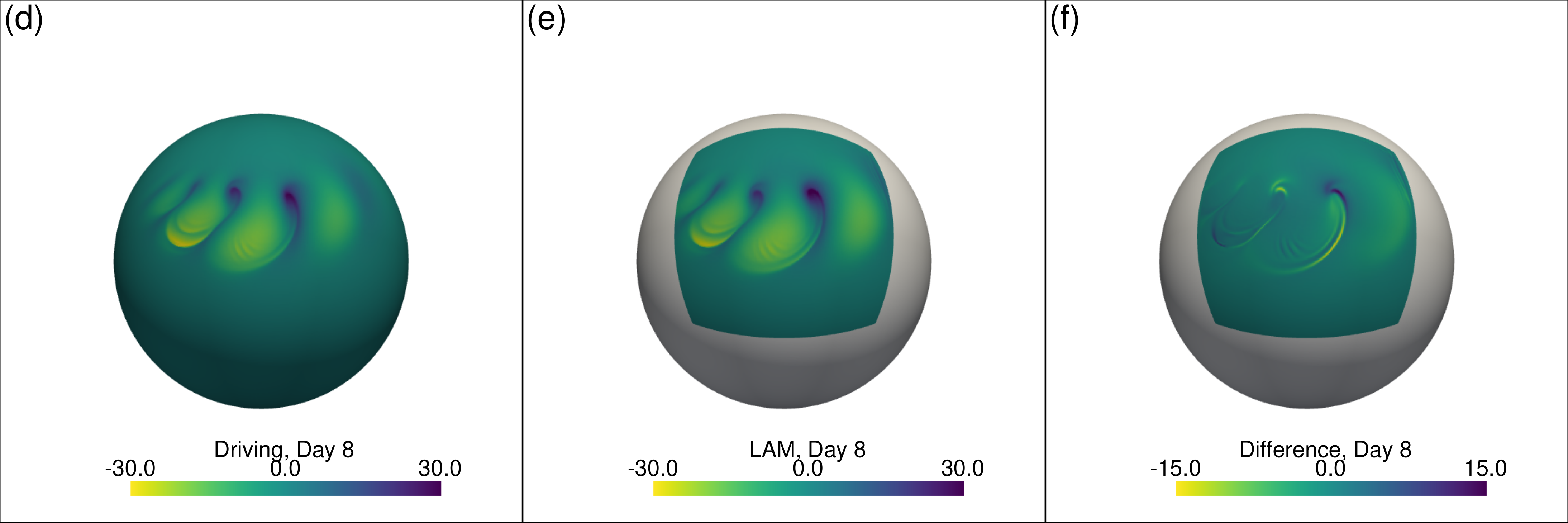}
\caption{{ Surface potential temperature $(^{\circ}C)$ with initial conditions subtracted after 8 days of simulation, using a C192 grid and the regional model using blending. Left: driving, Centre: regional, Right: regional minus driving.
{\label{fig:c192}}%
}}
\end{center}
\end{figure}

\begin{figure}[t!]
\begin{center}
\includegraphics[width=0.7\columnwidth]{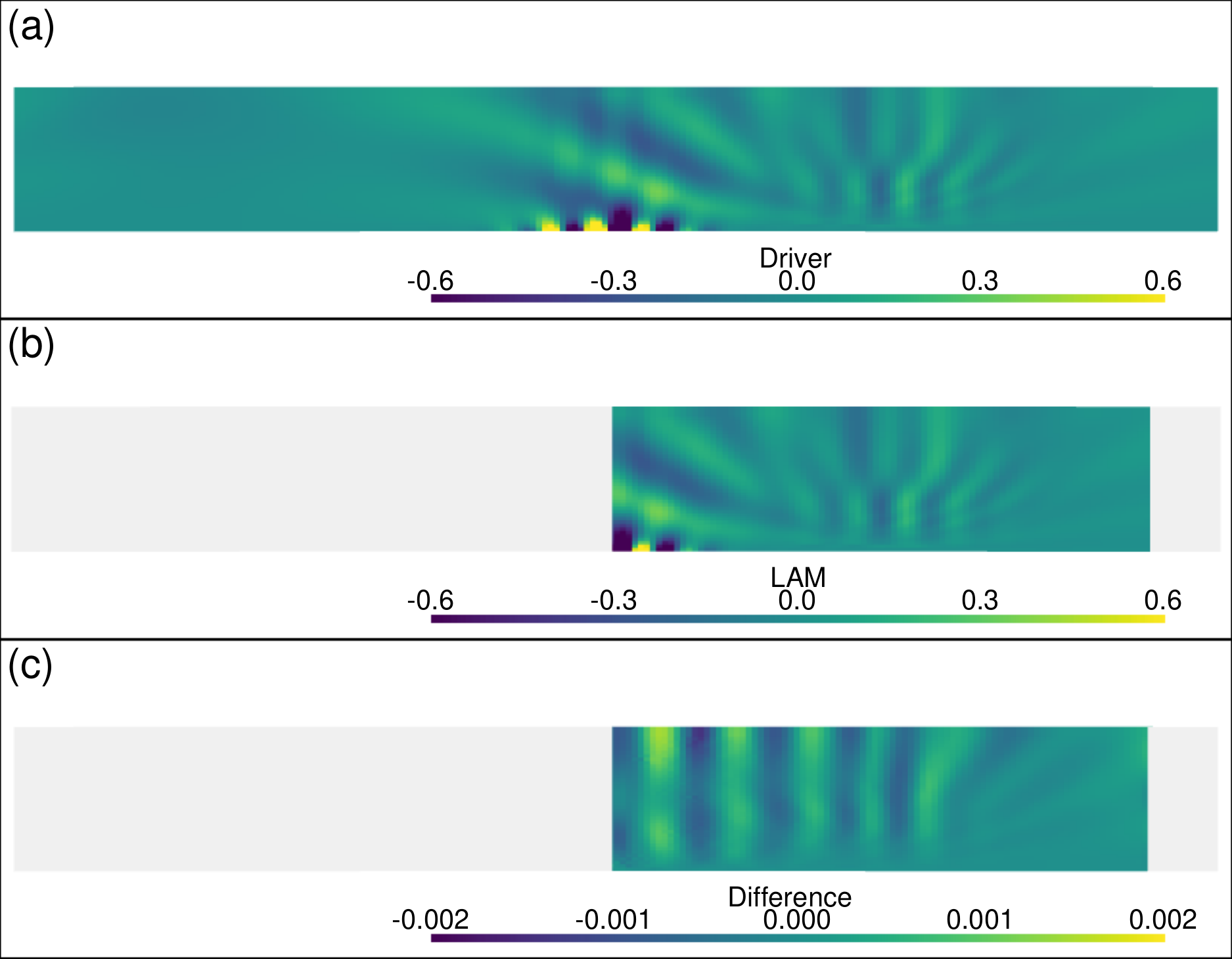}
\caption{Vertical velocity ($ms^{-1}$) for the 2D Sch\"{a}r hill test after 2400s (40mins), a) Driving, b) regional, c) regional minus driving. Only the lowest 40 levels (12km) of the model are shown.
{\label{fig:schar}}
}
\end{center}
\end{figure}

\subsubsection{Baroclinic wave: Higher resolution}
To demonstrate the applicability of these results to higher resolutions, the models are run with a C192 mesh, a timestep of 900s (15min) and the LBCs are provided at intervals of 48 timesteps (12 hours). The boundaries are positioned at a depth of 6 cells from the edge of the cubed-sphere panel, and blending is applied over a depth of 3 cells.

The results, after 8 days of simulation, are shown in Fig.~\ref{fig:c192}. A baroclinic wave has developed in the driving model, with a more detailed structure than for the C24 case, including a series of well-defined trailing fronts. The LAM simulation is able to recreate both the overall pattern and detail. The difference plot shows mainly a slight difference in the detail of the trailing fronts, with perhaps a slight phase error. But overall the results prove the formulation to be successful in application to higher resolution simulations.

\subsection{Sch\"{a}r Hill mountain wave tests}
The second test is based on the 2D Sch\"{a}r hill mountain wave test \cite[]{Schar:2002} case using a 2D Cartesian mesh with a 100km domain, 500m grid spacing, 100 vertical levels and 30km top, with a  damping layer applied in the top 10km, and a 20s timestep, similar to Fig.~6b in \cite{Melvin:2019}. The regional model is run on a section of the mesh with boundaries at the centre of the domain and at 5km from the right-hand edge, and with an LBC cell depth of 5 cells.

\subsubsection{Sch\"{a}r Hill: LBCs supplied every timestep}
The first experiment uses LBC data provided by the driving model on every timestep and blending is not applied. The vertical velocity is shown in Fig.~\ref{fig:schar}. In the driving model, the mountain range has triggered standing gravity waves that give rise to tilted layers with an alternating sign in the vertical velocity. The regional model is able to replicate this, with the result looking the same by eye.

\subsubsection{Sch\"{a}r Hill: Time-interpolated LBCs and blending}
\begin{figure}[t!]
\begin{center}
\includegraphics[width=0.7\columnwidth]{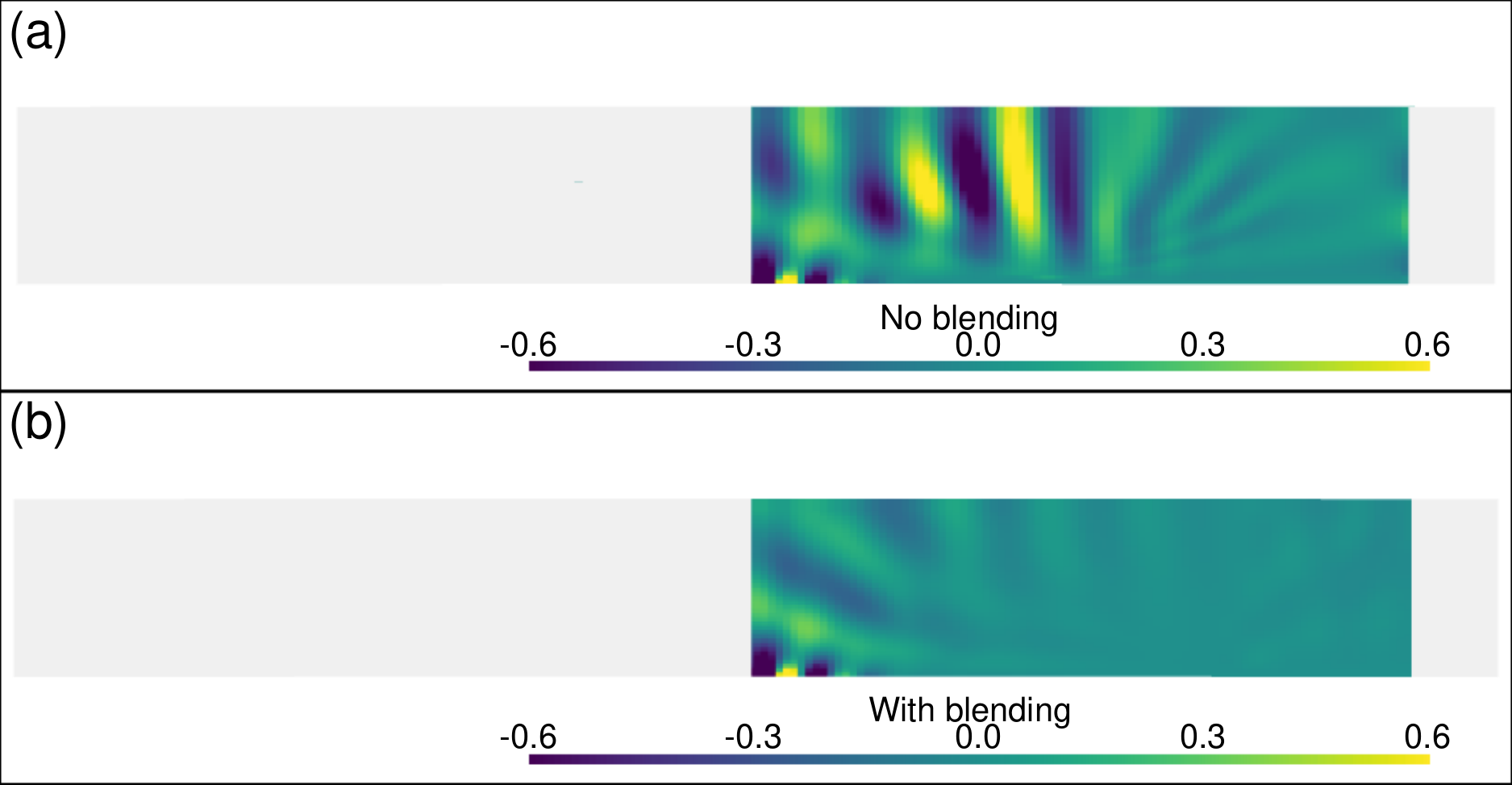}
\caption{As Fig.~\ref{fig:schar} but using less frequent LBC data and a) without blending, b) with blending.
{\label{fig:schar_lbc}}%
}
\end{center}
\end{figure}

The last experiment uses LBC data that is provided by the driving model every 100 timesteps, and a linear time interpolation is used for the intermediate timesteps. The regional model results are shown in Fig.~\ref{fig:schar_lbc}. These can be compared with the driving model results in  Fig.~\ref{fig:schar}a. Without blending (Fig.~\ref{fig:schar_lbc}a), the amplitude of the vertical velocity is much larger than in the driving model. This can likely be attributed to the inconsistency between the interior and the LBCs that then causes gravity waves to effectively become trapped in the interior, and not able to dissipate.

The experiment is repeated but with blending applied using linearly-varying weights over 25 cells depth. (It was found by numerical experimention that a depth of 3 had little impact, and that a larger depth such as 25 is required for this scenario). When blending is applied  (Fig.~\ref{fig:schar_lbc}b), the amplitude returns to a similar level as the driving model and the pattern of the waves is very similar to that of the driving model, close to the orography and the inflow boundary. Thus the blending allows the regional model to evolve the correct flow and then dissipate this close to the outflow boundary, so that large imbalances are not generated near to the outflow. In addition it demonstrates that the interior of the model can generate the correct small scale features as a result of the orography forcing, without needing this information to come through the LBCs.

\section{Conclusions}
We have developed a new formulation for a regional version of a mixed finite-element, semi-implicit dynamical core by adapting the global model equations to include lateral boundary conditions. This has been approached using the philosophy of only updating the model fields within a subdomain such that the solution is the same as would be obtained if the model fields were updated over the whole domain.

At the heart of the dynamical core is the Helmholtz-equation, in mixed system-from. We emulate solving the inhomogeneous boundary value problem by rearranging the linear equation so that the boundary values become part of the right-hand-side forcing.  We emulate the use of trial-functions with zero value on the boundary through the use of a pre- and post-multiplication by a zeroing-matrix which allows only the solution in the internal subdomain to be updated. These modifications then carry through to give a modified Helmholtz preconditioner. The choice of function spaces used by the mixed finite-elements in the dynamical core necessitates that the mixed-system uses the momentum/wind increment as the boundary condition, and this is equivalent to prescribing Neumann boundary conditions in the associated Helmholtz equation for pressure-only.

The regional dynamical core formulation has been tested using idealised dynamical core numerical tests, allowing for a perfect big brother experiment where the driving model and regional model use exactly the same resolution and numerical schemes. These experiments were based on a baroclinic wave test on the cubed sphere, and the Sch\"{a}r hill mountain wave test on a 2D periodic, Cartesian plane. With these choices, we have demonstrated that the regional model can be used with a variety of mesh structures. These numerical experiments have demonstrated the formulation of the regional dynamical core to work as expected.

A single iteration, single timestep baroclinic wave experiment has demonstrated the regional model result to be identical (within solver precision) to the driving model result. 

Long forecast experiments, with LBCs provided at every timestep, have shown the regional model results to be identical to the driving model results by eye. Similar experiments but with time-interpolated LBCs have shown the inconsistencies between the interior and the LBCs to be a significant problem, as anticipated. However, it was also demonstrated that these problems could be overcome through the use of applying blending, where the interior solution is gradually blended with the LBC data. This confirms the work of other studies that have shown blending to be an effective way of reducing the spurious impacts of imposing LBCs in regional models.

The work presented in this paper provides a step towards developing a full regional weather and climate model. The next main steps involve coupling the dynamical core with the physical parameterisations, creating a LAM mesh (e.g. a variable resolution \cite[]{Davies:2017}, rotated, lat lon mesh) and interpolating LBC data from the driving model to the LAM mesh. Operators at the edges of the mesh generally don't need to be adjusted; they might only perform one half of the actual calculation but these values will then be overwritten. These next steps have been undertaken at the Met Office, and have been successful in running full regional weather and climate forecasts. This will be presented in future papers.

\bibliographystyle{IEEEtran}
\bibliography{IEEEabrv,bib.bib}

\section{Appendix: Technical implementation using the LFRic infrastructure}
One of the key philosophies of the LFRic infrastructure \cite[]{Adams:2018} is the 'separation of concerns' which means that the science code is separated from the computer optimisation code. It does this through the use of an automatic code manipulation software known as PSyclone \cite[]{psyclone}.

The code is designed to loop over vertical columns, applying the science code to each column individually. However, this approach means that the science code does not have knowledge of the specific vertical column it is working on, necessitating the same code to be applied to all columns. This could pose challenges for the inclusion of lateral boundary conditions, which requires different equations to be solved in the columns at the edge of the mesh.

Here, the concept of binary masks (and zeroing-matrices) provides an efficient solution to this challenge. Once the binary mask has been defined, it can be applied to all columns simultaneously through the LFRic built-in kernels. For example, to calculate the \verb&interior& we can use the \verb&X_times_Y& function to multiply the \verb&full& field by the \verb&binary-mask&.
\begin{verbatim}
  call invoke( X_times_Y( &
       interior, full, binary-mask )
\end{verbatim}

There are various ways to calculate the binary masks. For the periodic meshes presented here, the binary mask is specified by comparing the physical coordinates with the specified boundary coordinates. For a non-periodic mesh, the binary mask can be specified using the mesh connectivity information. This involves querying whether a cell is connected to another cell on all sides. If a cell is not connected on all four sides, the cell value (initially set to zero) is incremented by 1. This process is repeated to check if the cell is connected to another cell with a value of 1 or more, and if so, the cell is incremented by 1. This process results in a field with layers around the edge of the domain, featuring gradually descending integers from the edge to the interior. This field can then be used to define the binary mask for boundaries specified at a particular cell-depth from the edge of the mesh.

The technique of using binary masks also allows flexibility for adopting future technical optimisations. For example, it might be that PSyclone is adjusted to only operate on a subset of columns and the masks could be used to determine the subset.

\section{Acknowledgements}
We would like to thank the many people that have contributed to the development of LFRic, PSyclone and GungHo and in particular to Thomas Allen and Terry Davies for their discussions on limited-area-modelling.

\section{Data availability statement}
The data used to support the findings of this study
were generated using the Met Office’s GungHo and LFRic-Atmosphere
models. The source code and configuration files are freely available from the \cite{MOSRS} upon
registration and completion of a software licence. The data
that support the findings of this study are available from
the corresponding author upon reasonable request. Scripts to visualise the data depend on the following open-source Python libraries: \cite{geovista}, \cite{iris}, \cite{matplotlib}.

\section{Author contributions}
C.Johnson and B. Shipway devised the conceptual ideas and formulation,
C. Johnson carried out the numerical simulations,
C. Johnson, T. Melvin and B. Shipway prepared the draft manuscript and designed the figures,
C. Johnson, B, Shipway, T. Melvin, T. Bendall, J. Kent, A. Brown, I. Boutle, M. Zerroukat and B. Buchenau contributed to the regional model numerical code,
B. Shipway and N. Wood managed the project,
and all authors reviewed the results and approved the final version of the manuscript.

\end{document}